\documentclass[11pt,twoside]{article}
\usepackage[activeacute,spanish]{babel}
\usepackage{baaa-cas}

\usepackage{graphicx}

\usepackage[T1]{fontenc} 

\usepackage{latexsym}
\usepackage{verbatim}

\begin{document}
\def\tablename{Table}%
\markboth{S.E. Nuza \& C. Scannapieco}%
{1st PD-BA Workshop on Massive Stars and Interacting Binaries - May 2024}

\pagestyle{myheadings}
%
%
\vspace*{0.3cm}
\parindent 0pt{Contributed talk}

\title{Cosmological gas accretion of Milky Way-type galaxies and the build-up of galactic discs}

\author{S.E. Nuza$^{1,2}$ \& C. Scannapieco$^{3,2}$}

\affil{%
  (1) Instituto de Astronom\'{\i}a y F\'{\i}sica del Espacio, CONICET–UBA, Argentina\\
  (2) Consejo Nacional de Investigaciones Cient\'{\i}ficas y T\'ecnicas, Argentina\\
  (3) Departamento de F\'{\i}sica, Facultad de Ciencias Exactas y Naturales, UBA, Argentina
}

\begin{abstract}
In this work, we present results on the assembly of stellar discs belonging to Milky Way-type galaxies in the Auriga simulated sample. We study the net accretion of gas onto the disc region as a function of time and radius to assess the feasibility of the so-called inside-out formation of galaxy discs. We found that most of the galaxies in our sample exhibit an inside-out disc growth, with younger stellar populations preferentially formed in the outer regions as accreted material turns into starts. This produces stable discs as long as late-time accretion is free from significant external perturbations. 
\end{abstract}

\section{Introduction}

Stellar discs in spiral galaxies form as a result of cosmological gas accretion and mergers through cosmic time. In this regard, the diversity of observed galactic morphologies naturally arises from the different merger and accretion histories of each dark matter halo, leading to the formation of different components such as the galactic bulge and the thin disc. Broadly speaking, bulge formation in the Milky Way (MW) is generally believed to occur over short timescales early in the Galaxy's lifetime, whereas the disc component is thought to form over more extended periods, sustained by gas accretion at increasingly larger radii, in a fashion that can be described from the ``inside-out'' (e.g. Nuza et al. 2019). 

In this contribution, we present results on the assembly of galactic discs from a sample of simulated MW-type galaxies from the Auriga project. In Section 2, we briefly present the simulated galaxy sample. In Section 3, we describe the methods used to compute the net gas accretion rate and timescales onto the stellar discs. In Section 4, we report our results. Finally, in Section 5, we present our conclusions. 

\section{The simulations}
\label{procedure}

The Auriga sample (Grand et al. 2017) comprises a set of 30 disc galaxies similar to the MW modelled within a full cosmological context with the cosmological code \textsc{arepo} (Springel et al. 2010). 
The cosmology adopted is consistent with the Planck collaboration XVI (2014) and the mass resolution of the model galaxies is about $3\times 10^5$ and $5\times 10^4\,\mathrm{M}_{\odot}$ for dark matter and gas, respectively. 

The galaxy formation model includes metal-line cooling, a uniform ultraviolet background for reionization, star formation, magnetic fields, and feedback from Type II/Ia supernovae, asymptotic giant branch stars and active galactic nuclei.  


The selection criteria required virial masses in the range of $1$–$2 \times 10^{12}\,\mathrm{M}_\odot$ and relative isolation, with no neighbouring halo with mass greater than $3\%$ of the galaxy’s mass within 9 times its virial radius. Further details on the simulated galaxies can be found in Grand et al. (2017) and Iza et al. (2022). 

\section{Methods}

Following Iza et al. (2022), we compute the net accretion rate of gas onto the stellar disc for each simulated galaxy in the sample. The net gain/loss of gas inside a ring contained within the plane of the disc is regulated by the amount of infalling and outflowing gas particles together with the instantaneous star formation rate, responsible for gas depletion. Then, the net gas accretion rate can be written as

\begin{equation}
    \dot{M}_{\rm net} = \frac{M_{\rm gas}(i) - M_{\rm gas}(i-1) + M_*}{t(i) - t(i - 1)}{\rm ,} 
\end{equation}

\noindent where $M_{\rm gas}$ is the gas mass, $i$ and $i-1$ are indexes labelling consecutive snapshots, $M_*$ is the stellar mass formed between them and $t$ is the cosmic time.


Moreover, we can define a {\it mass-weighted time} $\tau$ to characterise the timescale of net accretion onto each ring within the stellar disc, namely

\begin{equation}
    \tau\equiv\frac{\int_{t_i}^{t_0}t\dot{M}_{\rm net}(t){\rm d}t}{\int_{t_i}^{t_0}\dot{M}_{\rm net}(t){\rm d}t}
    {\rm ,}
\end{equation}

\noindent where $t_i$ is a reference time from which we start the calculation and $t_0$ is the age of the universe. As shown in Iza et al. (2022), stellar discs in the Auriga galaxies are well defined only after the first $\sim4\,$Gyr of evolution, hence, we adopt $t_i=4\,$Gyr (see also Iza et al. 2024).

\begin{figure}
\begin{center}
\hspace{-0.3cm}
\includegraphics[width=0.45\columnwidth]{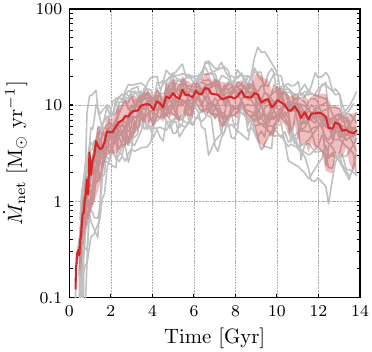}\hspace{0.2cm}\includegraphics[width=0.437\columnwidth]{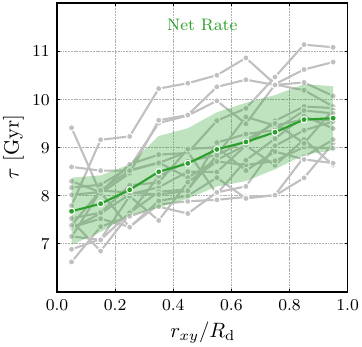}
\caption{{\it Left panel:} Net accretion rate of gas integrated onto the stellar discs as a function of cosmic time. {\it Right panel:} Net accretion timescale of gas integrated over time as a function of normalised disc radius. In both panels, grey solid lines correspond to individual galaxies in the Auriga sample. The mean and standard deviation in each case are also shown. Adapted from Iza et al. (2022, 2024).}
\label{fig1}
\end{center}
\end{figure}

\section{Results}
\label{results}

The left panel of Fig.~\ref{fig1} shows the net accretion rate of the Auriga galaxies as a function of cosmic time radially integrated onto the galaxy discs as defined in Iza et al. (2022). Despite differences between individual formation histories, the average net accretion rate shows a rapid increase at early times, reaching a maximum at $t\sim6\,$Gyr, to later display an exponential-like decline until the present epoch. 
The vast majority of galaxies show a moderate accretion during the last $\sim8\,$Gyr, giving rise to the formation of stable stellar discs.

The characteristic timescale of accretion as a function of normalised radius of the galaxies, integrated over their entire lifetime, can be seen in the right panel of Fig.~\ref{fig1}. As in the previous plot, the disc radius $R_{\rm d}$ for each individual galaxy is estimated following the calculations of Iza et al. (2022). In general, typical net accretion timescales increase with radius, implying a delayed accretion of material towards the external regions of the stellar discs. This behaviour is consistent with an inside-out formation of the discs, where older stars tend to be located in the inner galaxy regions, in contrast to younger stellar populations.  

\begin{figure}
\begin{center}
\hspace{-0.8cm}
\includegraphics[width=0.45\columnwidth]{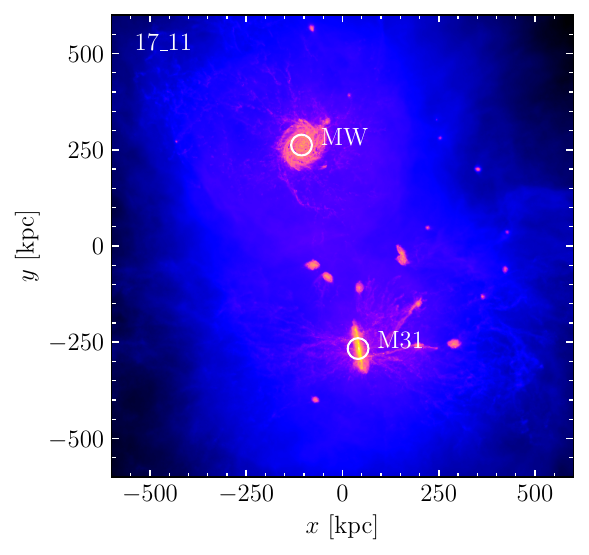}\hspace{0.2cm}\includegraphics[width=0.44\columnwidth]{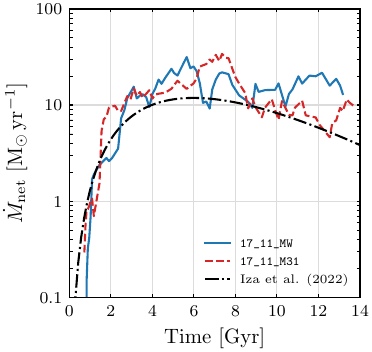}
\caption{{\it Left panel:} Gas distribution in the $17\_11$ {\sc hestia} realisation of the LG showing the MW and M31 galaxy candidates and their satellites. Solid circles indicate the virial radius. Adapted from Biaus et al. (2022). {\it Right panel:} Idem Fig.~\ref{fig1} (left panel) for MW and M31 candidates in {\sc hestia} compared to the best-fit curve of MW-like galaxies in Auriga given by Iza et al. (2022).}
\label{fig2}
\end{center}
\end{figure}

It is worth noting that the Auriga sample consists of a collection of relatively isolated MW-type galaxies drawn from a large cosmological volume. However, it is known that the MW galaxy is part of the Local Group (LG), a collection of galaxies dominated by the MW and Andromeda (M31), inhabiting an environment that may affect the evolution of the individual progenitors (e.g. Nuza et al. 2014; Creasey et al. 2015; Biaus et al. 2022). Preliminary results shown in Fig.~\ref{fig2} suggest that MW and M31 galaxies in one of the {\sc hestia} (Libeskind et al. 2020) LG simulations display larger net accretion rates than the overall Auriga sample. Further analyses on this matter will be performed in a future work.

\section{Conclusions}
\label{discussion}

In this work, we analysed a sample of simulated MW-type systems from the Auriga galaxy sample (Grand et al. 2017) to study the impact of gas accretion on the build-up of stellar discs as a function of cosmic time and radius. After computing the net accretion rate evolution of the sample and the radial dependence of the typical net accretion timescales in the simulated discs, we showed that model MW-type galaxies display a mean exponential-like behaviour with moderate rates required to build stable stellar discs during a significant fraction of the galaxy lifetimes. Additionally, preliminary results on a possible environmental dependence of gas accretion rates onto the MW-type systems were presented.

We also showed that the vast majority of galaxies in the sample display net gas accretion timescales that increase with disc radius suggesting that young stellar populations preferentially form in the external disc regions. These results, taken together, favour the inside-out formation scenario of stellar discs (e.g. Nuza et al. 2019; Iza et al. 2022, 2024).

\acknowledgements We would like to thank F.G.~Iza and L.~Biaus for adapting the figures presented in this contribution. S.E.N. and C.S. are members of the Carrera del Investigador Cient\'{\i}fico of CONICET. 

\begin{referencias}
    
\reference Biaus, L., Nuza, S.~E., Richter, P., et al., 2022, \mnras, 517, 6170
\reference Creasey, P., Scannapieco, C., Nuza, S.~E., et al., 2015, \apj, 800, L4
\reference Grand, R.~J.~J., G\'omez, F.~A., Marinacci, F., et al., \mnras, 467, 179
\reference Iza, F.~G., Scannapieco, C., Nuza, S.~E., et al., 2022, \mnras, 517, 832
\reference Iza, F.~G., Nuza, S.~E., Scannapieco, C., et al., 2024, \mnras, 528, 1737
\reference Libeskind, N.~I., Carlesi, E., Grand, R.~J.~J., et al., 2020, \mnras, 498, 2968
\reference Nuza, S.~E., Parisi, F., Scannapieco, C., et al., 2014, \mnras, 441, 2593
\reference Nuza, S.~E., Scannapieco, C., Chiappini, C., et al., 2019 \mnras, 482, 3089
\reference Planck collaboration XVI, 2014, \aap, 566, A54
\reference Springel, V., 2010, \mnras, 401, 791
\end{referencias}

\end{document}